\documentclass[twocolumn,showpacs,preprintnumbers,amsmath,amssymb]{revtex4}

\usepackage{graphicx}% Include figure files
\usepackage{dcolumn}% Align table columns on decimal point
\usepackage{bm}% bold math

\begin{document}

\title{Self-organized Boolean game on networks}

\author{Tao Zhou$^{1,2}$}
\author{Bing-Hong Wang$^{1}$}
\email{bhwang@ustc.edu.cn}
\author{Pei-Ling Zhou$^{2}$}
\author{Chun-Xia Yang$^{2}$}
\author{Jun Liu$^{2}$}
\affiliation{$^{1}$Department of Modern Physics, University of
Science and Technology of China, Hefei Anhui, 230026, PR China\\
$^{2}$Department of Electronic Science and Technology, University
of Science and Technology of China, Hefei Anhui,
230026, PR China\\
}

\date{\today}

\begin{abstract}
A model of Boolean game with only one free parameter $p$ that
denotes the strength of herd behavior is proposed where each agent
acts according to the information obtained from his neighbors in
network and those in the minority are rewarded. The simulation
results indicate that the dynamic of system is sensitive to
network topology, where the network of larger degree variance,
i.e. the system of greater information heterogeneity, leads to
less system profit. The system can self-organize to a stable state
and perform better than random choice game, although only the
local information is available to the agents. In addition, in
heterogeneity networks, the agents with more information gain more
than those with less information for a wide extent of herd
strength $p$.
\end{abstract}


\maketitle

\section{Introduction}
Complex adaptive systems composed of agents under mutual influence
have attracted considerable interest in recent years. It is not
unexpected that the systems with globally shared information can
be organized. A basic question in studies of complexity is how
large systems with only local information available to the agents
may become complex through a self-organized dynamical
process\cite{Paczuski2000}.

The mutual influence can be properly described as the so-called
information network, in which the nodes represent agents and the
directed edge from $x$ to $y$ means the agent $y$ can obtain
information from $x$. For simplicity, the undirected networks are
considered in this paper. In this way, node degree $k$ is
proportional to the quantity of information available to the
corresponding agent. The two extensively studied information
networks of ecosystem are
regular\cite{Moelbert2002,Zhou2004,Chau2004} and
random\cite{Paczuski2000,Galstyan2002} networks, both of which
have a characterized degree-the mean degree $\langle k\rangle$:
for regular networks, all the node are of degree $\langle
k\rangle$; and for random ones, the degree distribution decays
quickly in a Possionian form when $k>\langle k\rangle$. The
existence of characterized degree means every node has almost the
same capacity of information. However, previous empirical studies
have revealed that the information networks may be of scale-free
property\cite{Albert2002,Kullmann2002,Anghel2004}, in which the
giant heterogeneity of information exists. The nodes of larger
degree predominate much more information than those of less degree
thus the information heterogeneity can be measured by the degree
variance $\langle k^2\rangle$. The question is how the topology
affects the system dynamic, will the greater information
heterogeneity induce more profit for the system, or contraryly?

\begin{figure}
\scalebox{0.9}[1]{\includegraphics{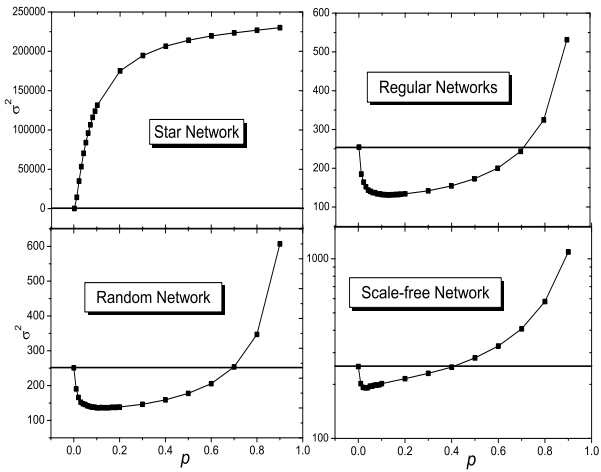}}
\caption{\label{fig:epsart} The variance of the number of agent
choosing +1 as a function of herd strength $p$. The four plots are
the cases of star, regular, random and scale-free networks,
respectively. The solid line represents the random choice game
where $\sigma^2=0.25N$. It is clear that the system profit is more
than random choice game when $p\in (0,0.7)$, $p\in (0,0.7)$ and
$p\in (0,0.4)$ in regular, random and scale-free networks,
respectively. For any $p\in (0,1)$, $\sigma^2$ of the four cases
satisfy that
$\sigma^2_{\texttt{regular}}<\sigma^2_{\texttt{random}}<\sigma^2_{\texttt{scale-free}}<\sigma^2_{\texttt{star}}$,
that means the system profit $S$ satisfy that
$S_{\texttt{regular}}>S_{\texttt{random}}>S_{\texttt{scale-free}}>S_{\texttt{star}}$.}
\end{figure}

\begin{figure}
\scalebox{0.8}[0.8]{\includegraphics{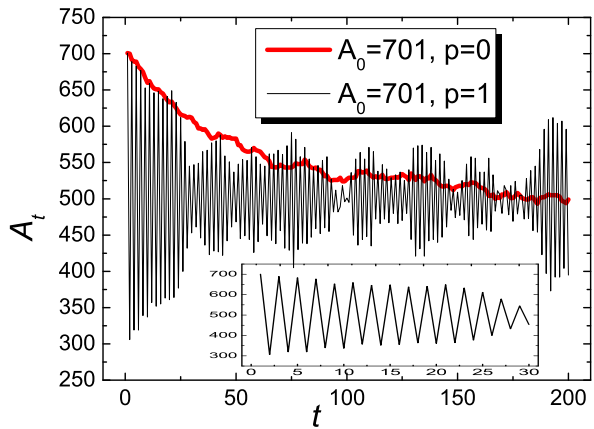}}
\caption{\label{fig:epsart} (Color online)The number of agent
choosing +1 vs time. The simulation takes place on regular
networks of size $N=1001$. At the beginning, a large event with
701 agent choosing +1 happens. The red thick and  black thin curve
show the variety of $A_t$ after this large event for the two
extreme cases $p=0$ and $p=1$, respectively. Clearly, in the case
$p=0$, $A_t$ slowly reverts to the equilibrium position $A\approx
\frac{N}{2}$; while in the case $p=1$, the system displays obvious
oscillation behavior. The inset exhibits the oscillation of $A_t$
in the case $p=1$ for the first 30 time steps.}
\end{figure}

Another question being concerned of in this paper is about the
herd behavior, which has been extensively studied in Behavioral
Finance and is usually considered as one factor of the origins of
complexity that may enhance the fluctuation and reduce the system
profit\cite{Eguiluz2000,Xie2005,Lee2004,Wang2005,Zhou2005}. Here
we argue that, to measure the strength of herd behavior, it is
more proper to look at how far the agents' actions are determined
by others rather than how far the agents want to be in majority,
since in many real-life cases, the agents would like to be in
minority but the herd behavior still occurs. We wonder whether
agents have different responses under a fixed herd strength, and
whether the variation trends of system profit and individual
profit are the same as the increase of herd strength.

In this paper, a model of Boolean game with only one free
parameter $p$ that denotes the strength of herd behavior is
proposed where each agent acts according to the information
obtained from his neighbors in network and those in the minority
are rewarded. Although the model may be too simple and rough, it
offers a starting point aiming at those questions above. We have
found that the topology of information network affects the system
dynamic much and the system can self-organize to a stable state
with more profit comparing with random choice game even only the
local information is available.

\section{Model}
Boolean game is firstly proposed by Kauffman where each agent has
only one binary choice such as either buying or selling a
stock\cite{Kauffman1993}. The studies of Boolean game have
attracted not only the physicists' but also the ecologists' and
economists' attention since it could explain many empirical data
and might contribute to the understanding of the underlying
mechanisms of the many-body ecosystems, although the dynamic rule
is simple\cite{Paczuski2000,Galstyan2002,Chen2004}.

Inspired by the idea of minority game\cite{Challet1997}, which is
a simple but rich model describing a population of selfish
individuals fighting for a common resource, we propose the present
Boolean game where each agent chooses between two opposing
actions, simplified as +1 and -1, and the agents in the minority
are rewarded. Each winner's score increases by one thus the system
profit equals to the number of
winners\cite{Challet1998,Savit1999,Quan2002}. In our model, at
each time step, each agent acts based on his neighbors at
probability $p$, or acts all by himself at probability $1-p$. In
the former case, we assume each neighbor has the same force. Since
the arbitrary agent $x$ would like to be in the minority, he will
choose +1 at the probability $\frac{s_{-1}^x}{s_{-1}^x+s_{+1}^x}$,
or choose -1 at the probability
$\frac{s_{+1}^x}{s_{-1}^x+s_{+1}^x}$, where $s_{-1}^x$ and
$s_{+1}^x$ denote the number of $x$'s neighbors choosing -1 and +1
in the last time step, respectively. In the latter case, since
there is no information from others, the agent will simply
inherits his action in the last time step or chooses the opposite
action at a small probability $m$, named mutation probability. It
is worthwhile to emphasize that, the agents do not know who are
winners in previous steps since the global information is not
available, which is also one of the main differences from the
previous studies on minority game.

The real-life ecosystem often seems a black box to us: the outcome
may be observed, but the underlying mechanism is not eyeable. If
we see many agents display the same action, we say the herd
behavior occurs, although those agents might prefer to be in the
minority. In another point of view, if each agent acts all by
himself, there is no preferential choice for +1 and -1 so as no
herd behavior will occur. Therefore, if the herd behavior occurs,
the agents' actions must be at least partly based on the
information obtained from others. In this paper, the strength of
herd behavior is measured by how far the agents' actions are
determined by others thus we set $p$ as the herd strength. This
measurement is not quantitative since in some networks there
exists an interval belong which the positive $p$ does not lead to
herd behavior comparing with the random choice game, however, it
is the measurement for the underlying possibility of the
occurrence of herd behavior.

\section{Simulations}
In this paper, all the simulation results are the average of 100
realizations and for each realizations the time length is $T=10^4$
unless a special statement is addressed. The number of agents
$N=1001$ and mutation probability $m=0.01$ are fixed. Figure one
shows the variance
$\sigma^2=\frac{1}{T}\sum^T_{t=1}(A_t-\frac{N}{2})^2$ as a
function of $p$ in star, regular, random and scale-free networks,
where $A_t$ is the number of agents who choose +1 at time step
$t$. Clearly, the smaller $\sigma^2$ corresponds to the more
system profit, and for the completely random choice game,
$\sigma^2=0.25N$. The regular network is a one-dimension lattice
with periodic boundary conditions and coordination number
$z=3$\cite{Newman1999}, the random network is the ER network of
connecting probability $6\times
10^{-3}$\cite{Erdos1960,Bollobas1985}, and the scale-free network
is the BA network of $m_0=m=3$\cite{Barabasi1999}. Therefore, all
the networks except the star networks are of average degree
$\langle k\rangle =6$. Since the number of edges $\frac{\langle
k\rangle N}{2}$ is proportional to the total quantity of
information available to agents, the networks used for simulating
except star networks have the same capacity of information. In
star network, it is not unexpected that the system profit will be
reduced when the herd strength increases. More interesting, in
each of the latter three cases, the system preforms better than
the random choice game when $p$ is in a certain interval,
indicating the self-organized process has taken place upon those
networks.

\begin{figure}
\scalebox{1}[1]{\includegraphics{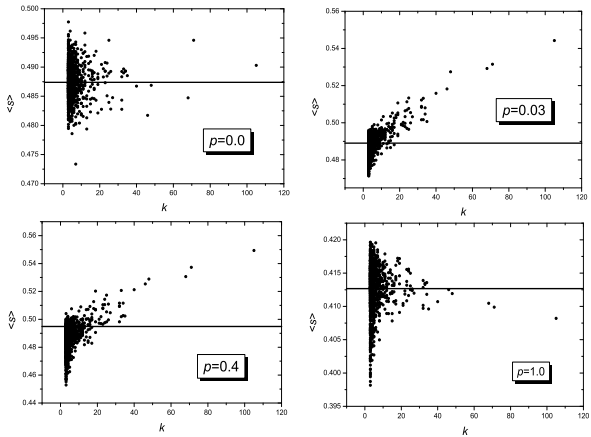}}
\caption{\label{fig:epsart} The agent's winning rate vs degree.
Each point denotes one agent and the solid line represents the
average winning rate over all the agents. In the cases of $p=0.0$
and $p=1.0$, no correlation is detected. In the cases of $p=0.03$
and $p=0.4$, the positive correlation between agent's profit and
degree is observed.}
\end{figure}
\begin{figure}
\scalebox{0.8}[0.9]{\includegraphics{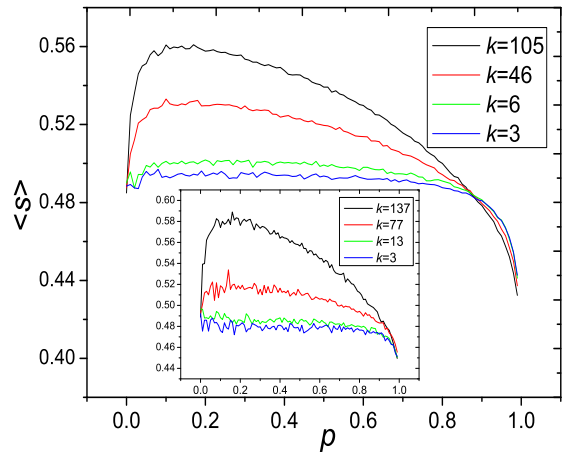}}
\caption{\label{fig:epsart} (Color online) The agent's winning
rate as a function of herd strength. The main plot is obtained by
the simulation upon a BA network of size $N=1001$, in which the
black, red, green and blue curves from up to bottom represent the
four agents of degree 105, 46, 6 and 3 respectively. The inset
shows the case upon a BA network of size $N=2001$, where the
black, red, green and blue curves from up to bottom represent the
four agents of degree 137, 77, 13 and 3 respectively. It is
observed that the agents having more information gain more than
those with less information.}
\end{figure}
Although having the same capacity of information, the dynamic of
scale-free networks is obviously distinguishable from that in
regular and random networks, indicating that the topology affects
the dynamic behavior much. Note that, although the topology of
regular and random networks are obviously different for they have
completely different average distance and clustering coefficient
and so on\cite{Watts1998}, the dynamic behaviors are almost the
same in those two networks. The common ground is they have almost
the same degree variance $\langle k^2 \rangle$. According to the
inequality $$\langle k^2 \rangle_{\texttt{star}}>\langle k^2
\rangle_{\texttt{scale-free}}>\langle k^2
\rangle_{\texttt{random}}>\langle k^2 \rangle_{\texttt{regular}}$$
and the simulation results, we suspect that the larger degree
variance, i.e. the greater information heterogeneity, will lead to
less system profit.

In figure one, one can see clearly that for all the four cases,
the variance $\sigma^2$ is remarkably greater than the random
choice game at large $p$. Consider the extreme case $p=1$, if the
agent choosing +1 and -1 are equably mixed up in the networks, and
the number of agent choosing +1 at present time is $A_t$, then in
the next time step, the expectation of $A_{t+1}$ is $\langle
A_{t+1} \rangle=N-A_t$, with departure $|\langle A_{t+1}
\rangle-\frac{N}{2}|=|A_t-\frac{N}{2}|$. If at present time $A_t$
is larger than $\frac{N}{2}$, then $A_{t+1}$ will be smaller than
$\frac{N}{2}$ most probably, and the departure from $\frac{N}{2}$
will not be reduced in average. Therefore, in the case of $p=1$,
when the ``large event" happens, that is to say $A_t$ is much
larger or much smaller than $\frac{N}{2}$ at some time $t$, there
will be a long duration of oscillation after $t$, in which $A$
skips between up-side $A>\frac{N}{2}$ and down-side
$A<\frac{N}{2}$. The oscillation behavior of $A$ is shown in
figure two. At the beginning, a large event with $A_0=701$ is
given, then the large oscillation goes on about 30 time steps. In
$p=1$ case, if $A$ gets apart from $\frac{N}{2}$, the influence
(large oscillation behavior) will stand for long time, leading
very large $\sigma^2$. However, in random choice game, whatever
$A_{t-1}$, the expectation of $A_t$ is always $\langle A_t
\rangle=\frac{N}{2}$, and the distribution of $A_t-\frac{N}{2}$
obeys Guassian form. That is the reason why the systems having
poor performance at large $p$ comparing with the random choice
game.

In another extreme case $p=0$, the expectation of $A_{t+1}$ is
$\langle A_{t+1} \rangle=A_t(1-m)+m(1-A_t)=A_t+m(1-2A_t)$. Assume
$A_t>\frac{N}{2}$, for $0<m<\frac{1}{2}$, we have
$A_t>A_{t+1}>\frac{N}{2}$. So in this case when $m$ close to zero,
no oscillation of $A$ will occur, but $A$ slowly reverts to the
equilibrium position $A\approx \frac{N}{2}$ after a large event.
One can easily prove that even for very small $m$, if the
iteration time $T$ is sufficiently long, the system profit will be
equal to random choice game, that means $\sigma^2=0.25N$. This is
strongly supported by the simulation results shown in figure one.
We also have check that the value of $m$ will not affect the
characters of these dynamic systems unless $m$ is very large. The
red thick curve in figure two is an example for the case $p=0$. At
time $t=0$, a large event $A_0=701$ occurs, and then the curve
$A_t$ slowly reverts to $\frac{N}{2}$. After about 170 time steps,
it arrives at the equilibrium position $A\approx \frac{N}{2}$.

The two extreme cases also exhibit a clear picture why the system
profit can be maximized at a special value of $p$. The herd
mechanism (with probability $p$) will bring oscillation, while the
independent mechanism (with probability $1-p$) will lead to a long
reversion process. The former mechanism makes $A$ skipping from
one side to another, while the latter one keeps $A$'s side. So,
under a proper value of $p$, the system can quickly arrive at the
equilibrium position $A\approx \frac{N}{2}$ after a large event
occurs, which leads to more system profit. The existence of
optimal $p$ has been demonstrated in figure one.

\begin{figure}
\scalebox{0.8}[0.8]{\includegraphics{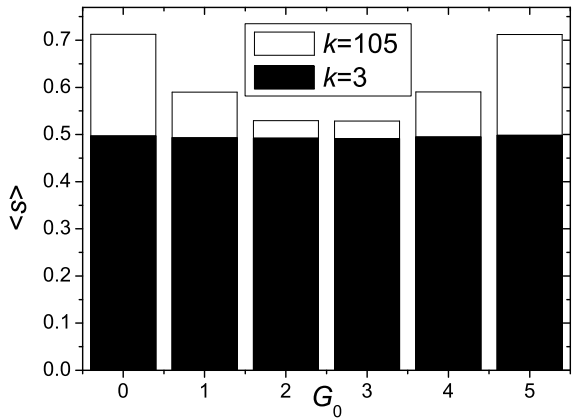}}
\caption{\label{fig:epsart} The agents' winning rate $\langle s
\rangle$ under different choice patterns $G_0$ of the five hub
nodes. The simulation takes place on the BA networks of size
$N=1001$, and the herd strength is fixed as $p=0.1$. The hollow
and solid histograms represent the winning rates of a hub node
($k=105$) and a small node ($k=3$), respectively. One can see
clearly, the winning rates of the small node under different
patterns are almost the same as $\langle s \rangle_{k=3}\approx
0.49$, which is obviously small than those of the hub node
especially in the case $G_0=0$ and $G_0=5$.}
\end{figure}
In succession, let's focus on the scale-free case since it may be
closer to reality. Firstly, we assume the agent choosing +1 and -1
are equably mixed up in the network. Since there is also no
degree-degree correlation for BA networks\cite{Newman2002}, for
arbitrary agent of degree $k$(here we do not differentiate between
node and the corresponding agent), the probability at which he
will choose +1 at time step $t+1$ is
\begin{widetext}
$$\eta_1(k,t+1)=p(\sum^k_{i=0}\frac{i}{k}C^i_k\rho^i_1(t)(1-\rho_1(t))^{k-i})+(1-p)(1-\rho_1(t))=1+2p\rho_1(t)-p-\rho_1(t),$$
\end{widetext}
where $\rho_1(t)$ denotes the density of agents choosing +1 at
time step $t$, and $C^i_k=\frac{k!}{(k-i)!i!}$. Since the
probability $\eta_1(k,t+1)$ is independent to $k$, there must be
no correlation between agent's degree and profit. In figure three,
we report the agent's winning rate vs degree, where the winning
rate is denoted by the average score $\langle s\rangle$ for
individual during one time step. $p=0.0$ and $p=1.0$ correspond to
the completely independent and dependent cases respectively,
$p=0.03$ is the point where the system performs best, and $p=0.4$
is another point where the system profit is equal to the random
choice game. One can see clearly that there exist the positive
correlation between agent's profit and degree in the cases
$p=0.03$ and $p=0.4$, which means the agents of larger degree will
perform better than those of less degree. Figure four shows the
agent's winning rate as a function of $p$ for different $k$. It is
clear that for a wide extent of $p$, the agents having more
information will gain more. Therefore, the assumption is not true,
thus there must be some kind of correlation, which is another
evidence of the existence of self-organized process.

A natural question is addressed: why the agents of large degree
will gain more than those of less degree? The reason is the choice
of a few hub nodes (i. e. the nodes of very large degree) can
strongly influence many other small nodes' (i. e. the nodes of
very small degree) choice in the next time step, and those hub
nodes can clean up from this influence. Denote $H$ the set of
those hub nodes and $G_0(t)$ the number of hun nodes choosing +1
at time $t$. We assume at a certain time step $t$,
$G_0(t)>\frac{|H|}{2}$, that means the number of hub nodes
choosing +1 is more than half. This departure will make some nodes
connecting to those hub nodes, especially the small nodes, choose
-1 in time $t+1$ with a greater probability. Because the majority
of these small nodes' hub neighbors choose +1 at present, this
influence is remarkable and can not be neglected since the small
nodes have only a few neighbors. The more departure
$|G_0-\frac{|H|}{2}|$ will lead to the greater influence.

Figure five exhibits an example on BA networks of size $N=1001$,
where $H$ contains only five hub nodes of the highest degree. In
each time step, all the choice of these five nodes form a choice
configuration. There are in total $2^5=32$ different
configurations, which are classified into 6 patterns by
identifying the number of agents choosing +1. For example, $G_0=2$
denotes the pattern that there are 2 agents choosing +1 and other
3 choosing -1. Under each choice pattern $0\leq G_0 \leq 5$, since
$|G_0-\frac{|H|}{2}|=|G_0-2.5|$ is bigger than zero at all time,
the hub node can always gain more than the small node. And
clearly, under the choice pattern with larger departure, such as
$G_0=0$ and $G_0=5$, the different of winning rates between the
hub node and the small node under these patterns is much greater
than the case of smaller departure.

\section{Conclusion}
In summary, inspired by the minority game, we propose a model of
Boolean game. the simulation results upon various networks is
shown, which indicate the dynamic of system is sensitive to the
topology of network, where the network of larger degree variance,
i.e. the system of greater information heterogeneity, leads to
less system profit. The system can perform better than the random
choice game. That is a bilievable evidence of the existence of
self-organized process taking place upon the networks although
only local information is available to agents. We also have found
that in heterogeneity networks, the agents with more information
gain more than those with less information for a wide extent of
herd strength $p$. In addition, it is clear that the trends of
varying of system profit and individual profit are different as
the increasing of herd strength, for example, in the scale-free
network with $p=0.5$, the system profit is less than random choice
game but the profit of agent of large degree is much more than
that in random choice game.

Although this model is rough, it offers a simple and
intuitionistic paradigm of the many-body systems that can
self-organize even when only local information is available. Since
the self-organized process is considered as one of the key
ingredients of the origins of complexity, it might contribute to
the understanding of the underlying mechanism of the complex
systems.

\begin{acknowledgements}
The authors wish to thank Mr. Bo Hu for his assistance in
preparing this manuscript. This work has been supported by the
National Science Foundation of China under Grant No.70171053,
70271070, 70471033 and 10472116; the Specialized Research Fund for
the Doctoral Program of Higher Education (SRFDP No.20020358009);
and the Foundation for graduate students of University of Science
and Technology of China under Grant No.KD200408.
\end{acknowledgements}

\end{document}